\documentclass[11pt]{article}
\usepackage{osid}
\usepackage{eepic,pstricks}

\usepackage{fancybox}

\usepackage[dvips]{graphicx}
\usepackage{amsmath,amssymb,mathrsfs}
\usepackage{epic,eepic}
\usepackage{epsfig}

\def\Tr{\operatorname{Tr}}

\def\>{\rangle}
\def\<{\langle}
\def\sH{\mathcal{H}}

\def\map#1{{\mathcal #1}}
\def\set#1{{\sf #1}}
\def\openone{I}
\def\alg#1{\mathscr #1}
\def\set#1{{\sf #1}}

\def\Base{\set{B}}

\def\dim{\operatorname{dim}}

\def\spc#1{\mathcal{#1}}

\def\Bndd#1,#2{\mathcal{B}(#1,#2)}

\def\Sex{S_\textrm{ex}}
\def\dual#1{#1^\prime}
\def\rank{\operatorname{rank}}

\begin{document}

\title{Quantum erasure of decoherence}

\author{Francesco Buscemi\\{\footnotesize\it Daini Hongo White Blgd. 201, 5-28-3 Hongo, Bunkyo-ku, 113-0033 Tokyo, Japan \& buscemi@qci.jst.go.jp}\\[2ex]
Giulio Chiribella and Giacomo Mauro D'Ariano\\{\footnotesize\it Dip.
di Fisica ``A. Volta'', via Bassi 6, I-27100 Pavia, Italy \&\\
chiribella@fisicavolta.unipv.it, ~~dariano@unipv.it} }

\maketitle
\begin{abstract}
We consider the classical algebra of observables that are diagonal in
a given orthonormal basis, and define a complete decoherence process
as a completely positive map that asymptotically converts any quantum
observable into a diagonal one, while preserving the elements of the
classical algebra. For quantum systems in dimension two and three any
decoherence process can be undone by collecting classical information
from the environment and using such an information to restore the
initial system state. As a relevant example, we illustrate the quantum
eraser of Scully \emph{et al.}  [Nature {\bf 351}, 111 (1991)] as an
example of environment-assisted correction.  Moreover, we present the
generalization of the eraser setup for $d-$dimensional systems,
showing that any von Neumann measurement on a system can be undone by
a complementary measurement on the environment.

\end{abstract}

\section{Introduction}

One of the fundamental postulates in Quantum Mechanics states that a
closed system naturally evolves according to a suitable unitary
transformation. It is then understood that every open system can in
principle be \emph{closed}, in the sense that, by extending the
boundaries of the subsystem of interest, it is in principle possible
to reach a situation in which everything inside the boundaries obeys a
global unitary evolution. In this case, \emph{information} is
conserved, that is, there is no net flow of information from the
global system. The global evolution preserves indeed the amount of
information that can be extracted from an arbitrary set of signal
states in which a classical alphabet is encoded, allowing only
transfers of information from a subsystem to another.

Here we are interested in a much more particular situation, that is,
when the quantum system of interest---the \emph{input}
system---unitarily interacts with an environment on which we can
perform measurements. In other words, even if the system itself
evolves as an open quantum system, according to the dynamics
described by a \emph{quantum channel}~\cite{kraus}, the complementary
subsystem closing the main system is bounded and can be monitored by
suitable measurements. We can then exploit a kind of
feedback control on the main system, in which we apply some opportune
corrections to the system, conditional on the outcomes of the
measurement that was performed on the environment. This procedure
is called \emph{environment-assisted channel correction}~\cite{GW} and
recently attracted  a lot of interest~\cite{interest}, also in
connection with the recently discovered \emph{state merging}
protocol~\cite{state-merg}.

In the present paper, we focus on a particular type of open system
dynamics, which are usually believed to play a fundamental role in
understanding the \emph{quantum-to-classical} transition, namely,
decohering evolutions~\cite{decoerenza}. This kind of channels causes
loss of coherence in quantum systems, and this phenomenon usually
constitutes the major practical limitation in quantum information processing. A large part of the literature concerning
quantum error correction is devoted to engineering methods to combat
the effects of decoherence~\cite{error-correction}. Here we propose a
decoherence correction method based on an environment-assisted
control, providing necessary and sufficient conditions for such a
method to be effective. Moreover, our analysis will be able to shed
some light on the information exchange dynamics between a quantum
system and the environment during a decohering evolution. From this
point of view, we will also review the quantum eraser
arrangement~\cite{quantum-eraser} as a particular example of
decohering evolution with a controllable environment, in which a
\emph{re-coherence} is possible conditional on the outcomes of a
suitable environment observable.

\section{Completely decohering evolutions}

Let's denote by $\alg A_q$ the ``quantum algebra'' of all bounded
operators on the Hilbert space $\spc H$, with $\dim\spc H=d<\infty$,
and by $\alg A_c$ the ``classical algebra'', namely any maximal
Abelian subalgebra $\alg A_c\subset\alg A_q$.  Clearly, all operators
in $\alg A_c$ can be jointly diagonalized on a common orthonormal
basis, which in the following will be denoted as $\Base =\{|k\>~|~k=1,
\dots ,d\}$.  Then, the classical algebra $\alg A_c$ is also the
linear span of the one-dimensional projectors $|k\>\<k|$, whence $\alg
A_c$ is a $d$-dimensional vector space. According to the above general
framework, we call \emph{(complete) decoherence map} a completely
positive identity-preserving (i. e.  trace-preserving in the
Schr\"odinger picture) map $\map E$ which asymptotically maps any
observable $O \in \alg A_q$ into a corresponding classical
observable  $O_c \in \alg A_c$, while preserving any element of the classical algebra $\alg A_c$. The defining properties of a decoherence map are then written explicitely as:
\begin{equation}\label{AsympClass}
\forall O \in \alg A_q: \qquad \exists \lim_{n \to \infty} \map E^n (O) \in \alg A_c 
\end{equation}
and
\begin{equation}\label{ClassPres}
\forall O_c \in \alg A_c: \qquad \map E (O_c) = O_c~.~ \qquad
\end{equation}

An important requirement in the above definition of decoherence
processes is that any classical observable is preserved. Notice that,
for example, the case of amplitude damping channels is not covered by
the definition, since in this case any state is driven to a fixed
state, namely not all classical observables are preserved.

It is easy to see that the set of decoherence maps is convex (i. e. if
we mix two decoherence maps we obtain again a decoherence map).
According to Eq. \eqref{ClassPres}, the set of decoherence maps is a
subset of the convex set of maps that preserve the elements of the
classical algebra $\alg A_c$. The convex structure of decoherence maps
has been analysed in Ref.~\cite{prl} using the following
representation theorem
\begin{theorem}{Theorem}
  A map $\map E$ preserves all elements of the classical
  algebra $\map A_c$ if and only if it has the form
\begin{equation}\label{SchurMap}
\map E(O)= \xi \circ O.
\end{equation}
$A \circ B$ denoting the Schur product of operators $A$ and $B$, i.~e.

\begin{equation} A \circ B \doteq \sum_{k,l=1}^d A_{kl} B_{kl} |k\>\<l|~,
\end{equation} 
$\{A_{kl}\}$ and $\{B_{kl}\}$ being the matrix elements of $A$ and $B$
in the basis $\Base$, and $\xi_{kl}$ being a correlation matrix,
i.e. a positive semidefinite matrix with $\xi_{kk}=1$ for all $k=1,
\dots ,d$.
\end{theorem}
Incidentally, notice that the operator $\xi$ in Eq.~(\ref{SchurMap})
is isometrically equivalent to the Choi operator~\cite{choi} $R_C =
\sum_{k,l}~\xi_{kl}~ |k\>|k\>\<l|\<l|$, which in turn is in one-to-one
linear correspondence with the Jamio\l kowski operator~\cite{jam}
$R_J= \sum_{k,l} ~\xi_{kl}~|l\>|k\> \<k|\<l|$.  Theorem~1 establishes
a one-to-one linear correspondence between maps that preserve the
classical algebra $\alg A_c$ and correlation matrices. This means that
both sets share exactly the same convex structure, whence a map is
extremal if and only if the corresponding correlation matrix is an
extreme point. The decohering evolutions, that have the additional
property of Eq. \eqref{AsympClass}, are represented by correlation
matrices $\xi$ with the property $|\xi_{kl}| < 1, \quad \forall k \not
= l$.

The extreme points of the set of correlation matrices have been
characterized by Li and Tam in Ref.~\cite{litam}. They proved that for
$d=2,3$, a correlation matrix is extremal if and only if it is
rank-one. This statement, translated in terms of maps, informs us
that, for $d=2,3$ extreme points of the convex set of maps that
preserve the classical algebra are  unitary
maps~\cite{prl}. As a consequence, for qubits and qutrits, every
decoherence map can be written as
\begin{equation}\label{RandomUnitary}
\map E(O)=\sum_ip_iU_i^\dag OU_i,
\end{equation}
where $U_i$'s are unitary operators and $p_i$ is a probability
distribution, namely any decoherence map is \emph{random-unitary}.
However, already for $d=4$ it is possible to explicitly
show~\cite{prl} that there exist extreme correlation matrices with
rank greater than one, and hence, decoherence maps that are not
random-unitary.

Notice that the action of the map $\map E$ in the Schr\"odinger
picture can be simply and \emph{uniquely} derived from the
trace-duality formula $\Tr[\map E(O)\ \rho]=\Tr[O\
\dual{\map E}(\rho)]$. From Eq.~(\ref{SchurMap}) it follows that, in
the case of decohering maps, $\dual{\map E}(\rho)=\xi^T\circ\rho$,
where $\xi^T$ denotes the transposition of the matrix $\xi$ with
respect to the fixed basis $\Base$ diagonalizing the classical
algebra. As a consequence, one has exponential decay of the
off-diagonal elements of $\rho$, since $|[\dual{\map E}{}^n
(\rho)]_{kl}|=|\xi_{lk}|^n\cdot |\rho_{kl}|$ and $|\xi_{kl}| <1 \quad \forall k \not = l$. In other words, any
initial state $\rho$ decays exponentially towards the completely
decohered state $\rho_\infty$ defined as 
\begin{equation}\label{rhoInfty}
\rho_{\infty}\doteq \sum_k\rho_{kk}|k\>\<k|~,
\end{equation}
namely, its diagonal with respect to the fixed basis $\Base$. Since a
matrix $\xi$ is a correlation matrix if and only if its transposition
$\xi^T$ is, in the following, when there is no possibility of
confusion, we will use the same symbol $\map E$ to denote the action
of the map on operators as well as on density matrices, also omitting
the transposition over $\xi$.

\section{Environment-assisted control}

In Ref.~\cite{GW}, the following general situation is considered. A
channel $\map E$, acting on density matrices $\rho$ on the input
Hilbert space $\sH$, is given. As a consequence of the Stinespring
theorem~\cite{stine}, we can always write it as follows~\cite{ozawa}
\begin{equation}\label{eq:unitary-real}
  \map E(\rho)=\Tr_e[U(\rho\otimes|0\>\<0|_e)U^\dag],
\end{equation}
namely, as a unitary interaction between the system and an
\emph{environment}, described by the Hilbert space $\sH_e$, followed
by a trace over the environment degrees of freedom. If the environment
input state is a pure one---like in
Eq.~(\ref{eq:unitary-real})---Gregoratti and Werner~\cite{GW} proved
that, assuming a somehow ``controllable'' environment, for all
possible unitary interactions $U$ in Eq.~(\ref{eq:unitary-real}), and
for all possible decompositions of the channel $\map E$ into pure
Kraus representations $\map E(\rho)=\sum_iE_i\rho
E_i^\dag$~\cite{kraus}, there exists a suitable rank-one POVM on the
environment, let us call it $\{|v_i\>\<v_i|_e\}$,
$\sum_i|v_i\>\<v_i|_e =\openone_e$, such that
\begin{equation}
  E_i\rho E_i^\dag=\Tr_e[U(\rho\otimes|0\>\<0|_e)U^\dag\ (\openone\otimes|v_i\>\<v_i|_e)].
\end{equation}
Within this setting, one can then think of performing a correction
$\map C_i$ on the system conditional on the $i$-th outcome of the
environment measurement, thus obtaining the following overall
corrected channel
\begin{equation}
\map E_\textrm{corr}(\rho)=\sum_i\map{C}_i(E_i\rho E_i^\dag).
\end{equation}
In Ref.~\cite{GW} it is shown that the only channels that can be
perfectly inverted by monitoring the environment---i.~e. such that it
is possible to have $\map E_\textrm{corr}(\rho)=\rho$, for all
$\rho$---are the random unitary ones. Therefore, it follows that one
can perfectly correct any decoherence map for qubits and qutrits by
monitoring the environment. The correction is achieved by retrieving
the index $i$ in Eq.~(\ref{RandomUnitary}) via the measurement on the
environment represented by the rank-one POVM $\{|v_i\>\<v_i|_e\}$, and
then by applying the inverse of the unitary transformation $U_i$ on
the system. Therefore, the random-unitary map simply leaks $H({p_i})$
bits of classical information into the environment, where $H$ denotes
the Shannon entropy and $p_i$ is the probability of the outcome
``$i$''. The effects of decoherence can be completely eliminated by
recovering such classical information, without any prior knowledge
about the input state.

\section{Bounds on the information flow}

It is now interesting to address the problem of estimating the amount
of classical information needed in order to invert a random-unitary
decoherence map. If the environment is initially in a pure state, say
$|0\>_e$, a useful quantity to deal with is the so-called entropy
exchange~\cite{Schumacher} $\Sex$ defined as
\begin{equation}\label{DefSex}
  \Sex(\rho)=S(\sigma_{e}^\rho),
\end{equation}
where $\sigma_{e}^\rho$ is the reduced environment state after the
interaction with the system in the state $\rho$, and
$S(\rho)=-\Tr[\rho\log\rho]$ is the von Neumann entropy.  The entropy exchange
quantifies the information flow from the system to the environment
and, for all input states $\rho$, one has the bound~\cite{Schumacher}
$|S(\map E(\rho))-S(\rho)|\le \Sex(\rho)$, namely the entropy exchange
$\Sex$ bounds the entropy production at each step of the decoherence
process.

In the case of initially pure environment, the entropy exchange
depends only on the map $\map E$ and on the input state of the system
$\rho$, regardless of the particular system-environment interaction
$U$ that is chosen to model $\map E$ via Eq. \eqref{eq:unitary-real}.
In particular, by the Kolmogorov decomposition for nonnegative
definite matrices it is always possible to write
$\xi_{kl}=\<e_l|e_k\>$ for a suitable set of normalized vectors
$\{|e_k\>\}$, and the map $\map E (\rho)=\xi\circ\rho$ can be realized
as $\map E(\rho)=\Tr_e[U(\rho\otimes |0\>\<0|_e)U^\dag]$, where the
unitary interaction $U$ gives the transformation
\begin{equation}\label{Unitary}
U|k\>\otimes|0\>_e=|k\>\otimes|e_k\>.
\end{equation}
With this choice of the interaction $U$, the final reduced state of
the environment is $\sigma_{e}^\rho=\sum_k\rho_{kk}|e_k\>\<e_k|$. In
order to evaluate the entropy exchange $\Sex$ for a decoherence map
$\map E(\rho)=\xi\circ\rho$, one can then use the formula
\begin{equation}\label{Sex_and_xi}
  \Sex(\rho)=S(\sqrt{\rho_\infty}\xi\sqrt{\rho_\infty}),
\end{equation}
which follows immediately from the fact that
$\sqrt{\rho_\infty}\xi\sqrt{\rho_\infty}$, and $\sigma_{e}^\rho$ are
both reduced states of the same bipartite state
$\sum_i\sqrt{\rho_{ii}}|i\>|e_i\>$.

When a map can be inverted by monitoring the environment---i.~e. in
the random-unitary case---the entropy exchange $\Sex(\openone/d)$
provides a lower bound to the amount of classical information $H({p_i})$
that must be collected from the environment in order to perform the
correction scheme of Ref.~\cite{GW}. In fact, assuming a
random-unitary decomposition (\ref{RandomUnitary}) and using the
formula~\cite{Schumacher} $\Sex (\rho)=S\left( \sum_{i,j} \sqrt{p_i
    p_j} \Tr[U_i \rho U_j^{\dag}] |i\>\<j| \right)$, we obtain
\begin{equation}\label{SexAndEntropy}
  \Sex(\openone/d) \leq H({p_i}).
\end{equation} 
The inequality comes from the fact that the diagonal entries of a
density matrix are always majorized by its eigenvalues, and it becomes
equality if and only if $\Tr[U_iU_j^{\dag}]/d=\delta_{ij}$, i.~e. the
map admits a random-unitary decomposition with \emph{orthogonal}
unitary operators. Moreover, from Eq. (\ref{Sex_and_xi}) we have $\Sex
(\openone/d)=S(\xi/d)$, whence the relation
\begin{equation}
H({p_i}) \ge  S (\xi /d)~, 
\end{equation}
which gives a lower bound on the amount of information needed from the
environment in order to invert the decohering evolution.

On the other hand, the random-unitary representation
(\ref{RandomUnitary}), when possible, is highly non unique. This means
that, depending on the particular unitary operators chosen, the
entropy $H({p_i})$ can be made as large as desired. However, it is still
possible to provide a (generally non tight) upper bound to the
\emph{minimum value} of the amount of classical information $H({p_i})$.
Such a bound is derived in Ref.~\cite{busc} as $H({p_i})\le
2\log\rank\xi$, and hence it generally holds that
\begin{equation}\label{Bounds}
  S(\xi/d)\le H({p_i})\le 2\log\rank\xi.
\end{equation}
Eq.~(\ref{Bounds}) is true for all dimensions $d$. It is then
reasonable that it does not accurately describe the peculiar geometry
enjoyed by two-dimensional systems. In fact, in Ref.~\cite{prl} it is
proved that for $d=2$, it always holds that
\begin{equation}\label{eq:qubits-info}
H({p_i})=S(\xi/2),\qquad d=2.
\end{equation}
However, already for $d=3$, there exist random-unitary decoherence
maps for which $S(\xi/d)<H({p_i})$ \emph{strictly}, and at the moment we
are not able to provide a better upper bound than the one given above.

\section{Example: the quantum eraser}

Our results about the possibility of inverting decohering evolutions
by collecting classical information from the environment can boast
a celebrated \emph{ante litteram} example, namely the quantum eraser
of Ref.~\cite{eraser}. In this Section we briefly review this example
using a compact notation that will turn out to be useful for its
generalization to similar cases in higher dimension.

\begin{figure}[h]
\begin{center}
\scalebox{0.53}{\input{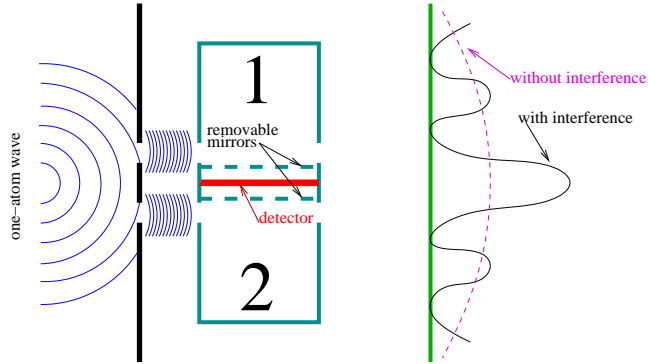}}
\caption{The Quantum Eraser arrangement.}
\label{fig:eraser}
\end{center}
\end{figure}

Let an excited atom pass through a double-slit, as depicted in
Fig~\ref{fig:eraser}.
Its state can be described in full generality by a density matrix
$\rho$, such that, if the orthogonal states $|1\>$ and $|2\>$
correspond to the particle passing through the slit number 1 or number
2, respectively, the probability of detecting the particle passing
through the slit number 1~(2) is $p(1)=\<1|\rho|1\>$
($p(2)=\<2|\rho|2\>$). Notice that $\rho$ can be a pure state, as in
the original quantum eraser proposal $\rho=|+\>\<+|$, with
$|+\>=1/\sqrt{2}~(|1\>+|2\>)$. 

If nothing is in between the slits and the collecting screen at the
end, fringes can be observed in the interference pattern, coming from
the non-null off-diagonal terms $\<1|\rho|2\>$ and $\<2|\rho|1\>$. But
if we place a probe, as in Fig.~\ref{fig:eraser}, consisting of two
resonant cavities initialized in the vacuum state $|0\>_p$, then
interference fringes disappear, since the atom, while relaxing to its
ground state, leaves a photon in one out of the two cavities, depending of
the slit it passed through.  The interaction of the atom with the probe can be described  by means of a controlled unitary $U$
of the form~(\ref{Unitary}), namely
\begin{equation}\label{eq:interprobe}
  U |i\>\otimes|0\>_p =|i\>\otimes|i\>_p,\qquad i=1,2,
\end{equation}
where $|1\>_p$ and $|2\>_p$ are the orthogonal states of the
electromagnetic field corresponding to the situations ``one photon in
cavity 1'' and ``one photon in cavity 2'', respectively. Since
$|1\>_p$ and $|2\>_p$ are orthogonal, the input state $\rho$
instantaneously collapses to its decohered final state $\rho_\infty$,
and off-diagonal terms are annihilated. This fact is usually
interpreted as saying that the probe, by means of the
interaction~(\ref{eq:interprobe}), keeps track of the \emph{which-way}
information about the atom's path, in such a way that such an
information can be in principle extracted by the experimenter.
Nevertheless, it is still possible to \emph{erase} the which-path from
the probe by measuring on it the Fourier-conjugate observable
$\{|+\>\<+|,|-\>\<-|\}$, where $|-\>=1/\sqrt{2}~ (|1\>-|2\>)$.
Experimentally, this can be done long \emph{after} the atom passed
through the cavities, by removing at once both mirrors in
Fig.~\ref{fig:eraser}, in such a way that the detector between the two
cavities is coupled with the symmetric state of the radiation inside
them~\cite{eraser}. Then, separating the two subensembles of events
corresponding to the measurement outcomes $+$ and $-$, it is possible
to retrieve the original interference fringes.

We interpret the whole double-slit setup as being a realization of
a completely decohering process described by the channel
\begin{equation}
  \map E (\rho)=\sum_{i=1}^2|i\>\<i|\rho|i\>\<i|=\openone\circ\rho.
\end{equation}
Such a channel is actually random-unitary (it is a decohering process
in dimension two), and hence is correctable by an
environment-assisted control procedure. In particular, for the
atom-radiation interaction given by Eq. \eqref{eq:interprobe}, by
measuring the observable $\{|+\>\<+|,|-\>\<-|\}$ on the probe, we
obtain a realization of the random-unitary Kraus representation
\begin{equation}
 \map E(\rho)=\frac 12\rho+\frac 12\sigma_z\rho\sigma_z,
\end{equation}
In conclusion, conditionally on the probe outcomes, both atom final
states conserve the original off-diagonal terms (a part of an
innocuous unitary rotation), and fringes appear on the intereference
pattern on the screen. Moreover, from Eq.~(\ref{eq:qubits-info}),
since $\xi =\openone$, we know that the erasure process
picks up from the probe $S(\openone/2)=1$ bit of information.

The quantum eraser can be simply generalized to the case of
instantaneous decoherence of $d-$dimensional quantum system. This
situation can be thought of as a kind of ``$d$-slits'' interference
experiment, where an excited atom emits a photon in one out of $d$
possible cavities. Analogously to the two-dimensional situation, the
correlation matrix describing the intantaneous decoherence channel is
$\xi =\openone$, namely one has $\map E
(\rho) = \sum_{i =1}^d~ \<i|\rho|i\>~|i\>\<i| = I \circ \rho$. The
channel itself admits the random-unitary representation
\begin{equation}
 \map E=\frac 1d\sum_{j=1}^dZ_j\rho Z_j^\dag,\qquad
  Z_j=\sum_{k=1}^de^{2\pi i\frac{kj}{d}}|k\>\<k|~,
\end{equation}
where the unitary operators $Z_j$'s generalize $\openone$ and
$\sigma_z$ to the $d$-dimensional case. In this case, for the system-probe interaction given by $U |i\> \otimes |0\>_p = |i\> \otimes |i\>_p,~i= 1, \dots, d$, the terms of the
random-unitary decomposition can be isolated by measuring the probe
observable $\{|\tilde e_j\>\<\tilde e_j|\}$, where the vectors
$|\tilde e_j\>$ are the Fourier transform
\begin{equation}
  |\tilde e_j\>=\frac {1}{\sqrt d}\sum_k e^{2\pi
    i\frac{j k}{d}}|k\>
\end{equation}
of the elements of the decoherence basis $\Base$. Once the measurement
outcome ``$j$'' is known, it is enough to undo the unitary $Z_j$ to
retrieve any unknown initial state $\rho$. The amount of classical
information to be erased from the probe is then equal to $H({p_i})=\log
d$.

An equivalent  way of presenting the $d-$dimensional eraser is
by stating that \emph{any von Neumann measurement on a system can be erased
by its Fourier complementary measurement on the environment.} The
istantaneous decoherence $\map E (\rho)$ $=~\sum_i \<i| \rho|i\>~ |i\>\<i|$ can be indeed considered as the effect of the von Neumann measurement
of the observable $\{|i\>\<i|\}$, while the interaction $U: |i\>
\otimes |0\>_p \mapsto |i\>\otimes |i\>_p$ can be viewed as the
transfer of classical information from the system to a quantum register. On
the other hand, the Fourier-complementary measurement $\{|\tilde
e_j\>\<\tilde e_j\}$ allows one to extract from the classical register
the information needed to restore coherence in the system. Quite
naturally, this amount of information is exactly the same amount that
was stored into the register, maximized over all possible unknown states $\rho$, i.e. $\log d$.

\section*{Acknowledgments}
Stimulating and enjoyable discussions with K~\.Zyczkowski are
gratefully acknowledged. This work has been supported by Ministero
Italiano dell'Universit\`a e della Ricerca (MIUR) through PRIN 2005.
F~B acknowledges Japan Science and Technology Agency for support
through the ERATO-SORST Project on Quantum Computation and
Information.

\appendix


\begin{thebibliography}{99}
\bibitem{kraus} K~Kraus, \emph{States, Effects, and Operations:
    Fundamental Notions in Quantum Theory}, Lect. Notes Phys. {\bf
    190} (Springer-Verlag, Berlin, 1983).
\bibitem{GW} M~Gregoratti and R~F~Werner, J.~Mod.~Opt. {\bf 50}, 915
  (2003).
\bibitem{interest} P~Hayden and C~King, Quantum Inform. Comput. {\bf
    5}, 156 (2005); J~A~Smolin, F~Verstraete, and A~Winter, Phys.
  Rev. A {\bf 72}, 052317 (2005); A~Winter, quant-ph/0507045.
\bibitem{state-merg} M~Horodecki, J~Oppenheim, and A~Winter, Nature
  {\bf 436}, 673 (2005); M~Horodecki, J~Oppenheim, and A~Winter,
  quant-ph/0512247.
\bibitem{decoerenza} W~H~Zurek, Phys. Today {\bf 44}, 36 (1991);
  W~H~Zurek, Rev. Mod. Phys. {\bf 75}, 715 (2003); M~Schlosshauer,
  Rev. Mod.  Phys. {\bf 76}, 1267 (2004).
\bibitem{error-correction} P~Shor, Phys. Rev. A {\bf 52}, 2493 (1995);
  A~M~Steane, Phys. Rev. Lett. {\bf 77}, 793 (1996); P~Zanardi and
  M~Rasetti, Phys. Rev. Lett. {\bf 79}, 3306 (1998); D~A~Lidar et al.,
  Phys. Rev. Lett. {\bf 81}, 2594 (1998); E~Knill et al., Phys. Rev.
  Lett. {\bf 84}, 2525 (2000); A~Yu~Kitaev, Annals Phys. {\bf 303},
  2-30 (2003); J~Preskill, in \emph{Introduction to Quantum
    Computation}, ed. by H-K~Lo, S Popescu, and T Spiller (World
  Scientic, Singapore, 1998).
\bibitem{quantum-eraser} M~O~Scully and K~Dr\"uhl, Phys. Rev. A {\bf
    25}, 2208 (1982); M~O~Scully, B-G~Englert, and H~Walther, Nature
  {\bf 351}, 111 (1991); S~D\"urr, T~Nonn, and G~Rempe, Nature {\bf
    395}, 33 (1998).
\bibitem{prl} F~Buscemi, G~Chiribella, and G~M~D'Ariano, Phys. Rev.
  Lett. {\bf 95}, 090501 (2005).
\bibitem{choi} M-D~Choi, Lin. Alg. Appl. {\bf 10}, 285 (1975).
\bibitem{jam} A~Jamio\l{}kowski, Rep. Math. Phys. {\bf 3}, 275
  (1972). 
\bibitem{litam} C-K~Li and B-S~Tam, SIAM J. Matrix Anal. Appl.  {\bf
    15}, 903 (1994).
\bibitem{stine} W~F~Stinespring, Proc.~Am.~Math.~Soc. {\bf 6}, 211
  (1955).
\bibitem{ozawa} M~Ozawa, J.~Math.~Phys. {\bf 25}, 79 (1984).
\bibitem{Schumacher} B~Schumacher, Phys. Rev. A {\bf 54}, 2614 (1996).
\bibitem{busc} F~Buscemi, quant-ph/0607034.
\bibitem{eraser} M~O~Scully, B-G~Englert, and H~Walther, Nature {\bf
    351}, 111 (1991).
\end{thebibliography}
\end{document}